# Control and Local Measurement of the Spin Chemical Potential in a Magnetic Insulator

Chunhui Du[1,†], Toeno Van der Sar[1,†], Tony X. Zhou[1,2,†], Pramey Upadhyaya[3], Francesco Casola[1,4], Huiliang Zhang[1,4], Mehmet C. Onbasli[5], Caroline A. Ross[5], Ronald L. Walsworth[1,4], Yaroslav Tserkovnyak[3], Amir Yacoby[1,2,*]

[1]Department of Physics, Harvard University, 17 Oxford Street, Cambridge, Massachusetts 02138, USA.

[2]John A. Paulson School of Engineering and Applied Sciences, Harvard University, Cambridge, Massachusetts 02138, USA.

[3]Department of Physics and Astronomy, University of California, 475 Portola Plaza, Los Angeles, California 90095, USA.

[4]Harvard-Smithsonian Center for Astrophysics, 60 Garden Street, Cambridge, Massachusetts 02138, USA.

[5]Department of Materials Science and Engineering, Massachusetts Institute of Technology, 77 Massachusetts Avenue, Cambridge, Massachusetts 02139, USA.

*Correspondence to: yacoby@physics.harvard.edu
†These authors contributed equally to this work.

**Abstract:** The spin chemical potential characterizes the tendency of spins to diffuse. Probing the spin chemical potential could provide insight into materials such as magnetic insulators and spin liquids and aid optimization of spintronic devices. Here, we introduce single-spin magnetometry as a generic platform for non-perturbative, nanoscale characterization of spin chemical potentials. We use this platform to investigate magnons in a magnetic insulator, surprisingly finding that the magnon chemical potential can be efficiently controlled by driving the system's ferromagnetic resonance. We introduce a symmetry-based two-fluid theory describing the underlying magnon processes, realize the first experimental determination of the local thermomagnonic torque, and illustrate the detection sensitivity using electrically controlled spin injection. Our results open the way for nanoscale control and imaging of spin transport in mesoscopic spin systems.

Control and measurement of the chemical potential of a spin system can be used to explore phenomena ranging from quantum phase transitions (*1, 2*), to Bose-Einstein condensation (*3, 4*), and spin transport in gases and solid-state systems (*5-10*). In recent decades, a large scientific effort has focused on harnessing spin transport for low-dissipation information processing (*6, 11-13*). However, setting up a desired spin chemical potential remains a challenging procedure, requiring high-power parametric excitation (*3*) or strongly dissipating electrical currents (*12*). In addition, very few techniques can locally measure a spin chemical potential (*3, 14*). A technique that allows spatial characterization on the nanometer scale, particularly important for exploring materials with short spin diffusion lengths such as typical metals (*15, 16*), has thus far remained elusive.

Because the spin chemical potential is inherently related to spin fluctuations, it can be quantitatively determined by measuring the magnetic fields generated by these fluctuations. We demonstrate this principle using the excellent magnetic field sensitivity of the $S=1$ electron spin associated with the nitrogen-vacancy (NV) center in diamond (*17, 18*). We measure the chemical potential of magnons - the elementary spin excitations of magnetic materials (*11*) - in a 20 nm-thick film of the magnetic insulator yttrium iron garnet (YIG) on a ~100 nm length scale (Fig. 1A). Our measurements reveal that the magnon chemical potential can be effectively controlled by exciting the system's ferromagnetic resonance (FMR) (Fig. 1B-C). We analyze these results using a model describing the coupling between the coherent spin order parameter and the thermal magnon bath, providing the first experimental determination of the local thermomagnonic torque which goes beyond the conventional spin caloritronic phenomenology (*10, 19*).

Locally probing the weak magnetic fields generated by the fluctuations of a spin system requires nanometer proximity of a magnetic field sensor to the system. We ensure such proximity by positioning diamond nanobeams (*20*) that contain individually addressable NV centers onto the YIG surface (Fig. 1D-E). We use a scanning confocal microscope to optically locate the NV centers and address their spin state (*21, 22*). A photoluminescence image (Fig. 1D) provides an overview of the system, showing an NV center (NV₁) in a nanobeam that is located within a few micrometers from the gold and platinum striplines used to excite magnons in the YIG.

Magnons generate a characteristic magnetic-field spectrum reflecting the occupation of the magnon density of states. We probe this spectrum using the sensitivity of the NV spin relaxation rates $\Gamma_\pm$ to magnetic field fluctuations at the NV electron spin resonance (ESR) frequencies $\omega_\pm$ (Fig. 1B-C) (*23*). For a system at thermal equilibrium, these rates can be expressed as (*22*)

$$\Gamma_\pm(\mu) = n(\omega_\pm, \mu) \int D(\omega_\pm, \mathbf{k}) f(\mathbf{k}, d) d\mathbf{k} + \Gamma_\pm^0 \qquad (1)$$

Here, $n(\omega, \mu) = \frac{k_B T}{\hbar\omega - \mu}$ is the Rayleigh-Jeans distribution (which is the Bose-Einstein distribution in the high temperature limit appropriate for our room temperature measurements), μ is the chemical potential, $k_B$ is Boltzmann's constant, $T$ is the temperature, $\hbar$ is the reduced Planck's constant, $D(\omega, \mathbf{k})$ is the magnon spectral density, $\mathbf{k}$ is the magnon wavevector, $f(\mathbf{k}, d)$ is a transfer function describing the magnon-generated fields at the NV site, $d$ is the distance of the NV to the YIG, and $\Gamma_\pm^0$ is an offset relaxation rate that is independent of the magnon spectrum. From Eq. 1 it is clear that, when $\Gamma_\pm(\mu) \gg \Gamma_\pm^0$, the chemical potential can be extracted in a way that is independent of many details of both sensor and system: normalizing the relaxation rate measured at μ = 0 by the relaxation rate measured at μ > 0 yields

$$\mu = \hbar\omega_\pm(1 - \frac{\Gamma_\pm(0)}{\Gamma_\pm(\mu)}) \qquad (2)$$

As a first step in gaining confidence in this procedure, we probe the magnetic-noise spectrum of the YIG film in the absence of external drive fields. We measure the NV spin relaxation rates as a function of an external magnetic field $B_{\text{ext}}$ applied along the NV axis (*22*), finding excellent agreement with the model described by Eq. 1 (Fig. 2A). Qualitatively, the observed field dependence can be understood by noting the high density of magnons just above the FMR frequency, which induces a peak in the $m_s=0\leftrightarrow-1$ NV$_1$ relaxation rate when the corresponding ESR frequency crosses this region (Fig. 2B). A fit allows us to extract the distance of the NV to the YIG film (*22*).

Next, we study the magnetic noise generated by the system under the application of a microwave (MW) drive field of amplitude $B_{\text{AC}}$ (*22*), using the sensitivity of the NV photoluminescence to magnetic fields at the ESR frequencies. Figure 2C shows the photoluminescence of NV$_1$ as a function of the drive frequency and $B_{\text{ext}}$. The straight lines result from the expected decrease in NV fluorescence when the drive frequency matches one of the NV ESR frequencies. In addition, the fluorescence decreases when the MW excitation frequency matches the calculated FMR condition of the YIG film. This effect results from an FMR-induced increase in the magnon density and associated magnetic-field noise at the NV ESR frequencies (*22, 24, 25*). A line-cut at $B_{\text{ext}} = 14.4$ mT shows a typical FMR linewidth of 8 MHz we observe in these measurements (inset Fig. 2C).

If the magnon occupation under the application of an FMR drive field can be described by the Rayleigh-Jeans distribution, Eq. 2 allows us to extract the chemical potential μ by measuring the NV relaxation rates. We measure the power dependence of the relaxation rate $\Gamma_-$ of NV$_1$ at several values of the external magnetic field (Fig. 3A). We find that $\Gamma_-$ increases with drive power $B_{\text{AC}}^2$, consistent with the FMR-induced decrease of NV photoluminescence shown in Fig. 2C. Strikingly, we observe that $\Gamma_-$ saturates as a function of drive power. Moreover, we find that the corresponding chemical potential saturates exactly at the minimum of the magnon band set by the FMR (*22*) (Fig. 3B), which is the maximum allowed value for a boson system in thermal equilibrium (*19*). We confirm this behavior using two different NV centers over a broad range of magnetic fields (Fig. 3C), providing compelling evidence that the magnon density is described by a finite chemical potential in the spectral region probed in this measurement. We independently verify that temperature does not change in these measurements (*22*). Another striking feature of our data is the initial slow increase of chemical potential observed at small $B_{ext}$ and low drive power (Figs. 3B, D).

The build-up of chemical potential under the application of an FMR drive field can be understood as a pumping process of thermal magnons by the FMR-induced precession of the coherent spin order parameter $s\mathbf{n}$, where $\mathbf{n}$ is a unit vector: an incoming thermal magnon scatters off the time-dependent $\mathbf{n}$, generating two thermal magnons and transferring one unit of angular momentum from the coherent spin density to the incoherent spin density $\tilde{n}$ (*22*). This process is the Onsager reciprocal of a local thermomagnonic-torque-induced precession of $\mathbf{n}$ (*19*), which is gaining increased attention in the field of spin caloritronics. Using a two-fluid phenomenology (*8*), we can describe the mutual dynamics of $\mathbf{n}$ and $\tilde{n}$ according to the following hydrodynamic equation (*22*)

$$\dot{\tilde{n}} = -\frac{\tilde{\alpha} s \mu}{\hbar^2} + \eta \cos\theta_n \frac{s}{\hbar} \mathbf{z} \cdot (\mathbf{n} \times \dot{\mathbf{n}}) \qquad (3)$$

Here, the first term on the right-hand side describes the decay of thermal magnons into the lattice, with $\tilde{\alpha}$ a constant related to Gilbert damping. The second term describes the pumping of thermal magnons by the FMR-induced precession of $\mathbf{n}$, with $\eta$ parametrizing the local thermomagnonic torque between $\mathbf{n}$ and $\tilde{n}$, and $\theta_n$ the instantaneous angle of $\mathbf{n}$ with respect to the sample-plane normal. By setting $\dot{\tilde{n}} = 0$ and averaging Eq. 3 over time under the assumption that the precession frequency of **n** is much faster than the pumping-induced change in magnon population, also known as the non-adiabatic pumping regime (*22*), we obtain

$$\mu = \kappa \frac{\eta}{\tilde{\alpha}} B_{AC}^2 \cos^2 \theta_{n_0} \quad (4)$$

where $\theta_{n_0}$ is the average magnetization angle with respect to the sample-plane normal, and κ is a parameter resulting from averaging over the elliptical motion of **n.**

A key prediction of this model, resulting from symmetry considerations (*22*), is that the coupling between **n** and $\tilde{n}$ vanishes for an in-plane orientation of the magnetization (i.e., for $\theta_{n_0} = \pi/2$). We can test this prediction using the measurement of the chemical potential as a function of $B_{ext}$ (Fig. 3D), as changing $B_{ext}$ changes $\theta_{n_0}$ in a well-defined way (*22*). We find that the dependence of the drive efficiency $d\mu/dB_{AC}^2$ on $B_{ext}$ in the low-power regime is accurately described by Eq. 4 (Fig. 3E). We highlight that the precise knowledge of the in-situ drive amplitude $B_{AC}$ provided by our NV sensor (*22*) is essential for this comparison. From a fit we extract $\eta \approx 10^{-4}$ (*22*). It is interesting to note that this value is comparable to the magnon-phonon coupling described by the YIG Gilbert damping parameter $\alpha \approx 10^{-4}$ (*26*), suggesting that thermal magnons can exert a torque large enough to induce a magnetization precession.

Finally, we illustrate the power of our technique by characterizing the chemical potential that results from electrically controlled spin injection via the spin Hall effect (SHE). The SHE is a phenomenon originating from spin-orbit interaction, in which a charge current generates a transverse spin current. Such a spin current can be injected into a magnetic system, a technique widely used to study non-equilibrium magnon properties (*13*, *15*, *27-29*). Figure 4A shows the measured relaxation rate of $NV_1$, located at ~1.7 μm from the edge of the Pt stripline (Fig.1D), as a function of the electrical current density $J_c$ in the Pt. We observe a clearly asymmetric dependence that is well described by a second-order polynomial (blue solid line)

$$\Gamma_-(\mu) = \Gamma_-(0) + \Gamma_1 + \Gamma_2 \qquad (5)$$

with $\Gamma_1 \propto J_c$ the linear part and $\Gamma_2 \propto J_c^2$ the quadratic part.

Intuitively, we may expect the quadratic part of this polynomial to result from heating due to Ohmic dissipation in the Pt wire and the linear part to result from the SHE. We check this expectation by exploiting the capability of the NV sensor to determine the temperature at the NV-site through measurements of changes in the zero-field splitting of the NV spin states (*22, 30*). We assume this temperature to be equal to the local YIG temperature because of the high thermal conductivity of diamond and the relatively insulating properties of air. We then use Eq. 1 to calculate the expected change in NV relaxation over the experimentally determined relevant temperature range of ~40 K (*22*). A comparison with the data shows excellent quantitative agreement (Fig. 4B), illustrating the unique potential of NV spins for probing heat-related magnon phenomena with applications in spin caloritronics such as studies of the spin Seebeck effect (*9, 10*).

We attribute the linear part of the current-induced change in relaxation rate to a change in chemical potential induced by the SHE. Importantly, we rule out a possible influence of the 'Oersted' field $B_{Pt}$ generated by the current in the Pt. Our NV sensor provides an in-situ measurement of $B_{Pt}$, allowing a control measurement as a function of an externally applied field that mimics $B_{Pt}$ (*22*). We do not discern a significant effect of such field on the NV relaxation rate (Fig. 4C). To extract the spin-Hall induced chemical potential from $\Gamma_1$, we expand Eq. 1 in the limit $\mu \ll \hbar\omega$ and assume a linear dependence $\mu \propto J_c$ (*13, 22*). We find that µ increases by ~0.1

GHz for $J_c = 1.2 \times 10^{11}$ A/m$^2$ (Fig. 4C), similar to values predicted by theory (*28*). Furthermore, we find that for a given current through the Pt, µ does not significantly depend on the spectral detuning between ω₋ and the FMR over a ~0.35 GHz frequency range, as determined by sweeping $B_{ext}$ (Fig. 4D). We note that in Fig. 4D, the spin-current injection efficiency is essentially constant as the magnetization angle varies by less than 0.6° (*22*).

Our results show that exciting the ferromagnetic resonance provides an efficient mechanism to control the magnon chemical potential. An analysis of this mechanism indicates that the local thermomagnonic torque in YIG is of the same order of magnitude as Gilbert damping. Confined magnon resonances such as edge modes in ferromagnetic strips could serve as local sources of spin chemical potential to control spin currents. The ability to measure spin chemical potentials with an ultimate imaging resolution set by the NV-sample distance opens up new possibilities for measuring spin density, currents, and conductance in mesoscopic spin systems, exploring diffusive and ballistic spin transport, and aiding the development of new spintronic nanodevices.

**Acknowledgments:**

We would like to thank Philip Kim for providing the setup for the nanobeam transfer, and Marc Warner and Michael Burek for help with nanobeam fabrication. This work is supported by the Gordon and Betty Moore Foundation's EPiQS Initiative through Grant GBMF4531. A.Y. and R.W. are also partly supported by the MURI QuISM project. Work at UCLA is supported by FAME, an SRC STARnet center sponsored by MARCO and DARPA, (P.U.) and U.S. Department of Energy, Office of Basic Energy Sciences under Award No. DE-SC0012190 (Y.T.). R.W. and H.Z. would like to thank the DARPA QuASAR program and the Smithsonian Institution. Work at MIT was supported by the Solid-State Solar-Thermal Energy Conversion Center (S3TEC), an Energy Frontier Research Center funded by DOE, Office of Science, BES under Award # DE-SC0001299 / DE-FG02-09ER46577. F.C. acknowledges support from the Swiss National Science Foundation (SNSF), grant No. P300P2-158417. Device fabrication was performed at the Center for Nanoscale Systems (CNS), a member of the National Nanotechnology Infrastructure Network (NNIN), which is supported by the National Science Foundation under NSF award no. ECS-0335765. CNS is part of Harvard University.

**Author contributions:**

C.D. and A.Y. conceived the idea and designed the project. C.D. led the project. C.D., T.S., and A.Y. designed the measurement schemes and analyzed the data. C.D. fabricated the devices and performed the measurements. T.Z. built the confocal setup and contributed to the device fabrication. P.U. and Y.T. developed the theory of FMR-pumping of thermal magnons. F.C. and H.Z. proposed and fabricated the nanobeams. M.O. and C.R. provided the YIG sample. C.D., T.S. and A.Y. wrote the manuscript with the help from all co-authors. C.D., T.S., T.Z., P.U., Y.T., F.C., H.Z., R.W. and A.Y. contributed to the discussions. A.Y. supervised the project.

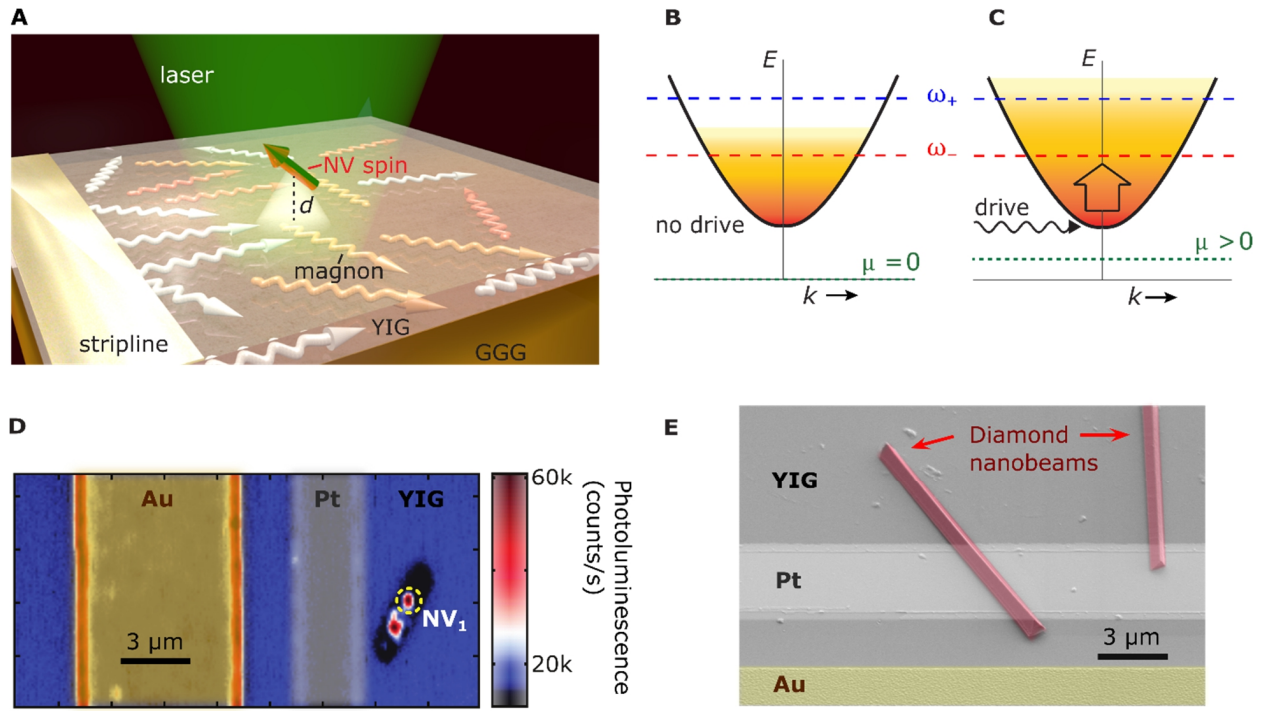

**Fig. 1. Local control and measurement of the magnon chemical potential.** (**A**) An NV spin locally probes the magnetic fields generated by magnons in a 20 nm-thick YIG film grown on a $Gd_3Ga_5O_{12}$ (GGG) substrate. (**B**) Sketch of the magnon dispersion and the magnon density at zero chemical potential, which falls off as 1/Energy (1/$E$) as indicated by the fading colors. (**C**) Driving at the ferromagnetic resonance increases the magnon chemical potential. The NV probes the magnon density at the electron spin resonance frequencies $\omega_{\pm}$. (**D**) Photoluminescence image showing a diamond nanobeam containing individually addressable nitrogen-vacancy sensor spins positioned on top of the YIG film. A 600 nm-thick Au stripline (false-colored yellow) provides microwave control of the magnon chemical potential and the NV spin states. A 10 nm-thick Pt stripline (false-colored grey) provides spin injection through the spin Hall effect. (**E**). Scanning electron microscope image of representative diamond nanobeams.

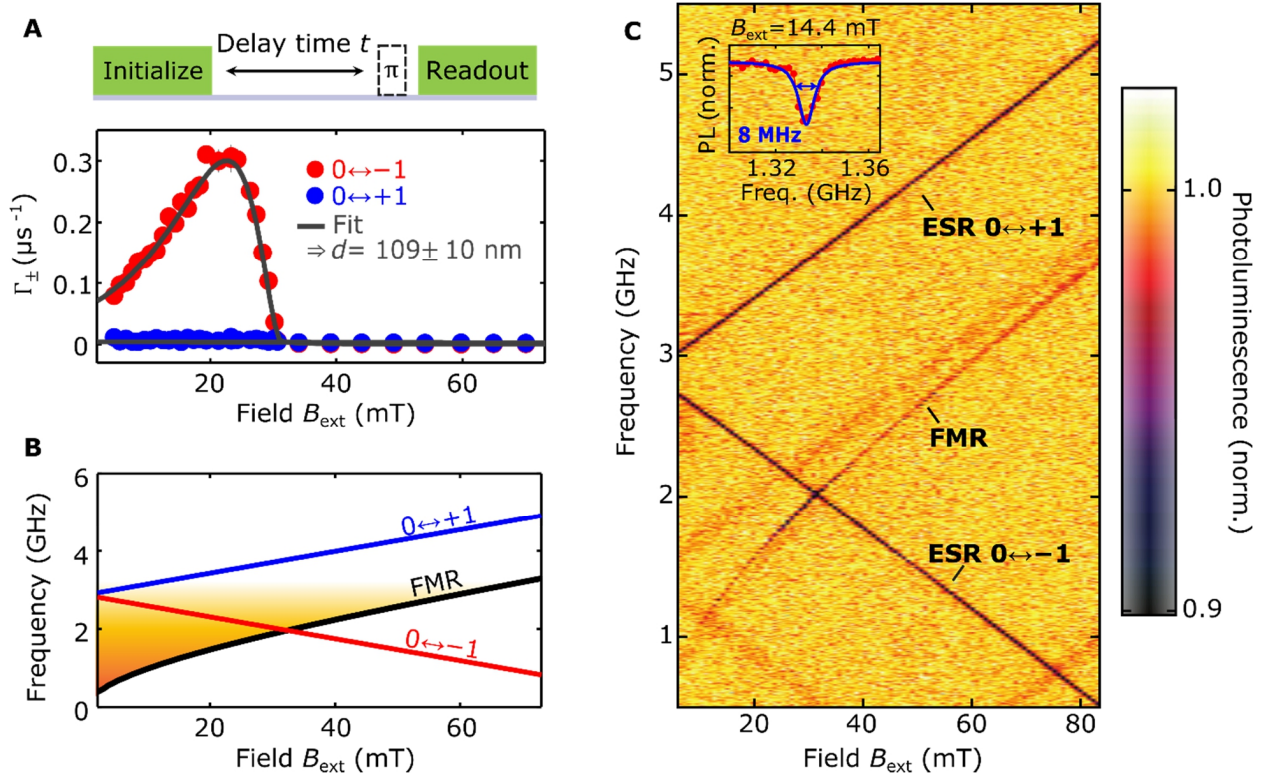

**Fig. 2. Tools for characterizing the magnon chemical potential. (A)** Measured spin relaxation rates $\Gamma_\pm$ corresponding to the $m_s=0\leftrightarrow\pm1$ transitions of NV$_1$ as a function of an external magnetic field $B_{ext}$. A fit to Eq. 1 yields the distance $d$ of NV$_1$ to the YIG film. The measurement sequence is depicted on top. Green laser pulses are used for spin initialization and readout. The three-level relaxation dynamics of the NV spin are characterized using MW pi pulses on the appropriate ESR transitions. $B_{ext}$ is applied along the axis of NV$_1$, at a $\theta = 65°$ angle with respect to the sample-plane normal, and a $\phi = 52°$ in-plane angle with respect to the Au stripline. **(B)** Sketch of the magnon density and the NV ESR frequencies vs $B_{ext}$. **(C)** Normalized photoluminescence (PL) of NV$_1$ as a function of $B_{ext}$ and the frequency of a 0.17 mT microwave drive field. The labeled (unlabeled) straight lines correspond to the NV ESR transitions in the electronic ground (excited) state (*18*). Inset: linecut showing the 8 MHz linewidth of the YIG ferromagnetic resonance (FMR) at $B_{ext}$=14.4 mT.

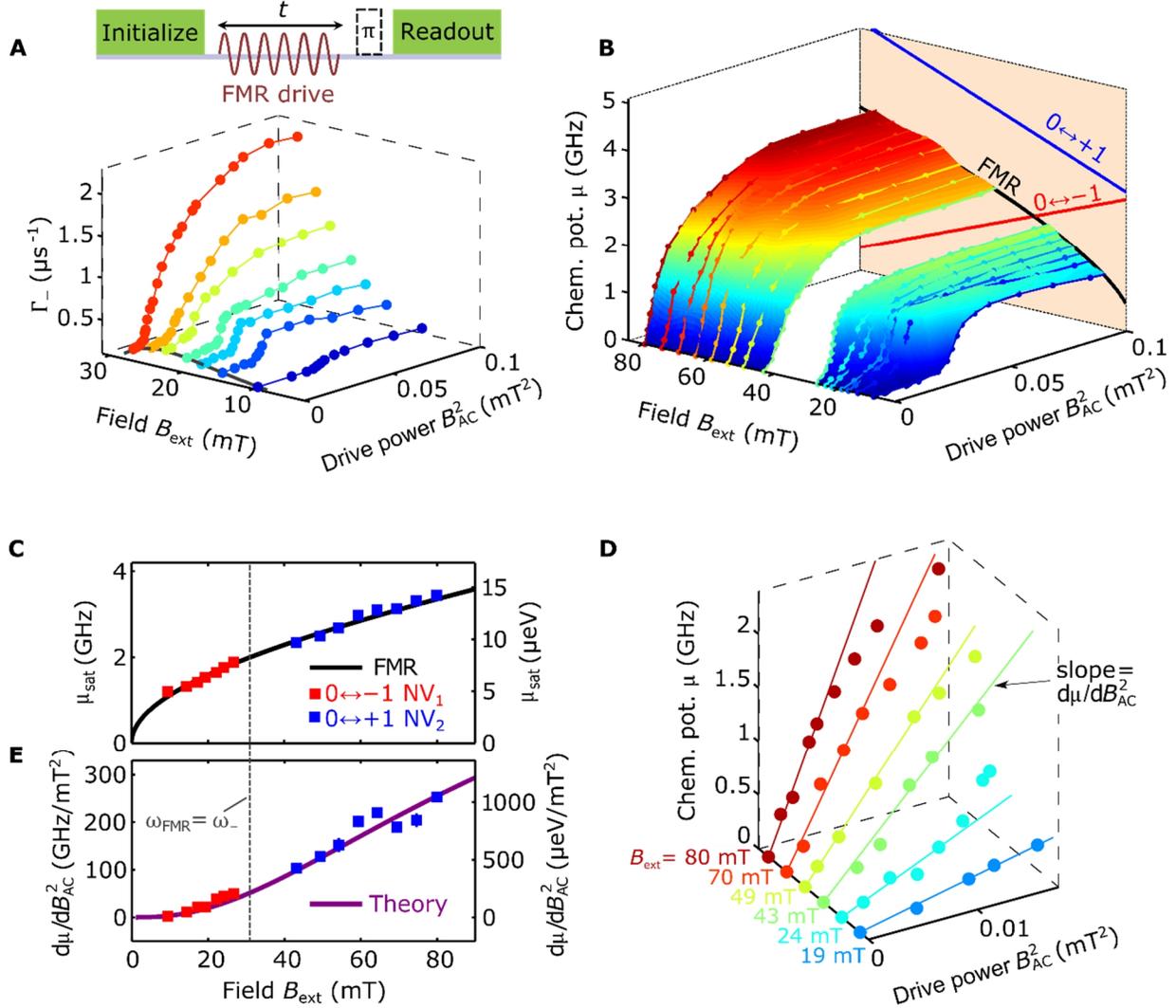

**Fig. 3. Magnon chemical potential (μ) under ferromagnetic resonance (FMR) excitation. (A)** NV₁ relaxation rate $\Gamma_-$ as a function of the on-chip power $B_{AC}^2$ of a magnetic drive field applied at the FMR frequency, for different values of $B_{ext}$. The grey curve is the fit from Fig. 2A. Top: measurement sequence. **(B)** μ as a function of $B_{AC}^2$ and $B_{ext}$. μ saturates at the minimum of the magnon band set by the FMR frequency. **(C)** Field dependence of the saturated value of the chemical potential $\mu_{sat}$ calculated from averaging μ in the region 0.05 mT² < $B_{AC}^2$ < 0.1 mT² (see B). The black curve is the FMR. The red (blue) points are measured using the NV₁ $m_s=0\leftrightarrow-1$ (NV₂ $m_s=0\leftrightarrow+1$) transition respectively. $B_{ext}$ is oriented along the NV axis: at a θ = 65° angle with respect to the sample-plane normal for both NVs, and a $\phi = 52°$ ($\phi = 6°$) in-plane angle with respect to the Au stripline for NV$_1$ (NV$_2$). The NV$_2$-YIG distance is 65±10 nm (*22*). **(D)** At low $B_{AC}^2$, µ increases linearly at a rate $d\mu/dB_{AC}^2$ that depends on $B_{ext}$. **(E)** Field dependence of $d\mu/dB_{AC}^2$ extracted from (D). A comparison to theory yields the local thermomagnonic torque (see text).

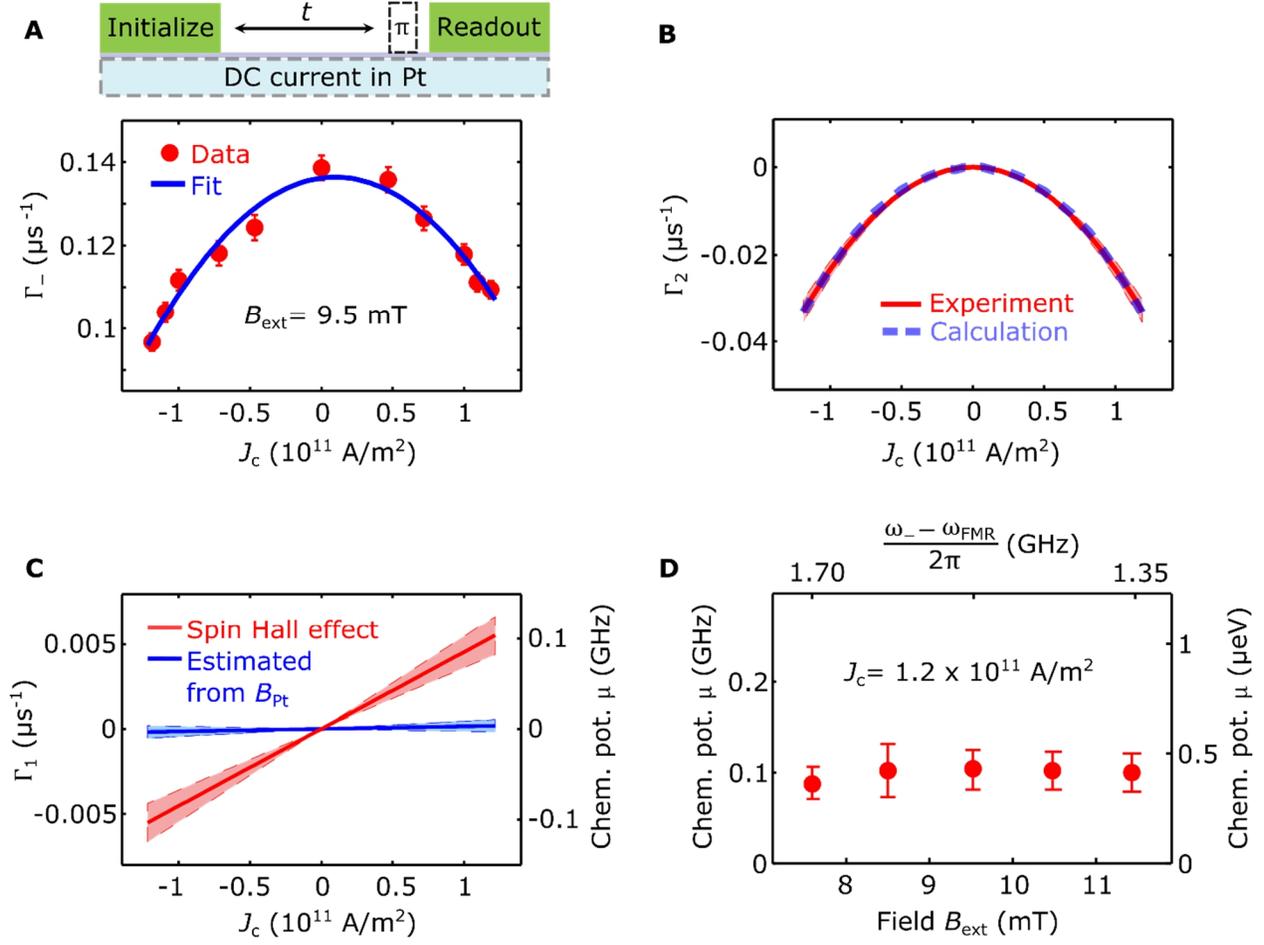

**Fig. 4. Magnon chemical potential resulting from the spin Hall effect (SHE). (A)** Measured NV relaxation rate $\Gamma_-$ vs current density $J_c$ in the Pt stripline. Blue curve: second-order polynomial fit. Top: measurement sequence. **(B)** Quadratic part of the measured change in NV relaxation rate, extracted from (A), compared to a calculation based on the experimentally determined increase in temperature resulting from Ohmic dissipation in the Pt wire. **(C)** Linear part of the measured change in NV relaxation rate (red line), attributed to the SHE, from which we extract the chemical potential as a function of $J_c$. A control measurement (*22*) shows that the contribution of the DC field $B_{Pt}$ generated by the current in the Pt is negligible (blue line). **(D)** Field dependence of the chemical potential. In (B) and (C), the shaded regions indicate 2 s.d. confidence intervals based on the uncertainty of the fit parameters.

18